\documentclass[12pt]{iopart} 

\usepackage{graphicx}
\usepackage{bm}
\usepackage{latexsym}
\usepackage{iopams}

\usepackage{color}

\begin{document}

\title{Inflation with Multi-Vector-Hair: The Fate of Anisotropy}

\author{Kei Yamamoto$^{1)}$, Masa-aki Watanabe$^{2)}$ and Jiro Soda$^{2)}$}
\address{1) DAMTP, Centre for Mathematical Sciences, University of Cambridge,
Wilberforce Raod, Cambridge CB3 0WA, United Kingdom \\
2) Department of Physics,  Kyoto University, Kyoto, 606-8502, 
Japan}
\eads{\mailto{K.Yamamoto@damtp.cam.ac.uk}, \mailto{jiro@tap.scphys.kyoto-u.ac.jp}}

\date{\today}

\begin{abstract}
We study inflation with multiple vector fields. 
In the presence of non-trivial couplings between the inflaton and the vector fields,
it turns out that no-hair conjecture does not hold and vector-hair appears.
In the case of uniform couplings, nevertheless, we find that
the universe approaches an isotropic final state after transient anisotropic inflationary phases.
For general couplings, we numerically show attractors are anisotropic inflation.
Even in these cases, it turns out that the inflation always tends to minimize the anisotropy
in the expansion of the universe. 
\end{abstract}

\maketitle

\section{Introduction}

It is believed that inflation accounts for features 
of cosmic microwave background radiation (CMB) observed by WMAP~\cite{Komatsu:2010fb}. 
The remarkable predictive power of inflation is associated with the cosmic
no-hair conjecture which claims that any classical hair will disappear
exponentially fast and the universe asymptotically approaches  de Sitter spacetime once the vacuum energy dominates the universe.
This conjecture is supported by the no-hair theorem which holds in the special situations~\cite{Wald:1983ky,varun}.
Thus, remaining quantum vacuum fluctuations are responsible for
 seeds of the large scale structure of the universe. 
The nature of quantum fluctuations reflects symmetry of de Sitter spacetime and gives rise to the following robust predictions. 
Firstly, the power spectrum of quantum fluctuations should have the statistical isotropy 
 due to the rotational symmetry of the spatial slices. 
 Secondly, the isometries of de Sitter result in constant Hubble expansion rate and the existence of the preferred Banch-Davis vacuum, which in turn implies the scale independent power spectrum. Finally, the slow roll conditions are guaranteed by shift symmetry in field space
which implies suppression of nonlinearity and hence Gaussian statistics of fluctuations. 
Indeed, these predictions have been observationally confirmed.

It should be stressed that the cosmic no-hair conjecture is just a conjecture.
 Historically, there have been challenges to the cosmic no-hair 
 conjecture~\cite{Ford:1989me,Kaloper:1991rw,Barrow:2005qv,Barrow:2009gx,Campanelli:2009tk,Golovnev:2008cf,Kanno:2008gn,Ackerman:2007nb}. 
Unfortunately, it turned out that these models 
suffer from either the instability~\cite{Himmetoglu:2008zp,Golovnev:2009rm,EspositoFarese:2009aj}, or a fine tuning problem, or a naturalness problem. 
Recently, however, stable anisotropic inflationary solutions in the context of supergravity have been found as
 counter examples to the cosmic no-hair conjecture~\cite{Watanabe:2009ct,Kanno:2009ei}.
 More precisely, it has been shown that, in the presence of a vector field coupled with the inflaton,
there could be a small anisotropy in the expansion rate which never
decays during inflation. 
Subsequently, the generality of anisotropic hair in inflationary universes has been also 
investigated~\cite{Kanno:2010nr,Moniz:2010cm,Hassan,Dimopoulos:2010xq,Do:2011zz,Murata:2011wv,Bhowmick:2011em,Hervik:2011xm}.

In spite of the failure of the cosmic no-hair conjecture, since the anisotropic inflation is an attractor
solution, the predictive power of the model still remains~\cite{Himmetoglu:2009mk,Dulaney:2010sq,Gumrukcuoglu:2010yc,Watanabe:2010fh}.
The interesting point is that the small anisotropy results in observationally relevant signatures.
In fact, there exist imprints of the anisotropic expansion in the CMB~\cite{Watanabe:2010bu}.
Therefore, it is worth studying to what extent the cosmic no-hair conjecture can be violated.
To this end, we examine inflationary models with multiple vector fields.
In the presence of non-trivial couplings between the inflaton and the vector fields, 
the energy density of vector fields never dies away, which should give rise to anisotropy
in the cosmic expansion. However, it has been known that multiple vector fields can take
a configuration for which the universe is nearly, or entirely isotropic \cite{Golovnev:2008cf}. 
So, the question is the fate of the cosmic anisotropy in the presence of non-vanishing vector fields.
It turns out that the accelerating universes tend to minimize their spatial anisotropy,  
which might be regarded as a generalization of cosmic no-hair conjecture, \i.e., the cosmic minimum-hair conjecture. 
For uniform gauge-kinetic coupling models, we find an isotropically inflating spacetime is the attractor whenever there are three or more vector fields.
Although the analysis is carried out for Abelian vector fields, it also suggests that non-Abelian gauge fields, which are described by multi-component vector fields, cannot be anisotropic in the asymptotic future if the inflation is eternal.
In more general cases, even though the anisotropy will not disappear, it will be minimized during inflation.

The organization of the paper is as follows.
In section II, we define models and give basic equations. 
In section III, we show the existence of isotropic inflationary solutions and
the stability of isotropic solutions. This highlights a tendency of isotropization of the inflating universe. 
In section IV, we examined the global structure of the phase space.
In particular, we investigate the fate of the anisotropy for general couplings. 
 The final section is devoted to a brief discussion on phenomenology and the conclusion.

\section{Models}
\label{sc:basic}

In this section, we setup models. For simplicity, we take power-law type inflationary models
for which we have exact solutions even in the presence of vector fields~\cite{Kanno:2010nr}.

We consider the following action.
\begin{equation}
\fl S = \int d^4 x \sqrt{-h}\left( \frac{1}{16\pi G}R-\frac{1}{2}\partial _{a }\phi \partial ^{a }\phi - V_0 e^{-k\phi }
-\frac{1}{4}\sum  _{m=1}^N e^{g_m \phi } F^{(m)}_{ab}F^{(m)ab} \right) ,
\end{equation}
where $R$ is the Ricci scalar calculated from the metric $h_{ab}$ and $G$ is the Newton constant.
The $N$ copies of Abelian gauge field $A_a^{(m)}$ with field strength $F_{ab}^{(m)} = \partial_a A_b^{(m)} -\partial_b A_a^{(m)} $ are
 coupled to an inflaton $\phi$ with coupling constants $g_m$. We consider the exponential potential with parameters $V_0$ and $k$. 
The Latin letters denote the full space-time indices. Greeks are reserved for spatial parts. 

Following  Ellis and MacCallum \cite{Ellis:1968vb}, we introduce the unit normal $u^a$ of the homogeneous hypersurface in generic Bianchi I space-times. The evolution of the spatial slice is described by
\begin{equation}
 u_{a;b} = H(h_{ab}+u_a u_b ) + \sigma _{ab} \ ,
\end{equation}
where $H$ is the averaged expansion rate and the trace-free tensor $\sigma _{ab}$ measures anisotropy of the universe. The decomposition of energy-momentum tensor can be carried out as
\begin{equation}
T_{ab}= \rho u_a u_b + 2q_{(a}u_{b)} + p (h_{ab}+u_a u_b ) + \pi _{ab} .
\end{equation}
$\rho ,p, q_a$ an $\pi _{ab}$ are respectively energy density, pressure, energy flux and anisotropic pressure seen by the observer with 4-velocity $u^a$. 
The electric fields seen by an observer with the four-velocity $u^a$ are defined by
\begin{equation}
E^{(m)}_a = F_{ab}^{(m)}u^b 
\end{equation}
and we assume, again for simplicity, that the magnetic parts vanish, that is
\begin{equation}
\eta _{abcd}F^{(m)bc}u^d = 0
\end{equation}
 where $\eta _{abcd}$ is the volume four-form. Then the matter variables are given as follows:
\begin{eqnarray}
\rho &=& \frac{1}{2}\dot{\phi }^2 +V_0 e^{-k \phi } + \frac{1}{2}\sum _{m=1}^N e^{g_m \phi } E_a^{(m)}E^{(m)a} \\
q_a &=& 0 \\
p &=& \frac{1}{2}\dot{\phi }^2 - V_0 e^{-k\phi } +\frac{1}{6}\sum _{m=1}^N e^{g_m \phi } E_a^{(m)}E^{(m)a} \\
\pi _{ab} &=& \sum _{m=1}^N e^{g_m \phi }\left(- E^{(m)}_a E^{(m)}_b + \frac{1}{3}E_c^{(m)}E^{(m)c}(h_{ab}+u_a u_b) \right) .
\end{eqnarray}
Here overdots denote differentiation with respect to the proper time $t$ associated with the unit normal $u^a$.
 
In a group invariant orthonormal frame, the Einstein equations take the form given in Wainwright and Ellis \cite{WE}. Let us assume that $E^{(1)}_{\alpha }$ and $E^{(2)}_{\alpha }$ are non-vanishing and not parallel to each other.  We choose the spatial frame such that 
\begin{equation}
E_{\alpha }^{(1)} = ( E,0,0) , \ \ \ \ \ E_3^{(2)}= 0.
\end{equation}
This fixes the gauge completely and results in
\begin{equation}
\Omega _1 = \sigma _{23} , \ \ \ \ \ \Omega _{2} =-\sigma _{13} , \ \ \ \ \ \Omega _3 = \sigma _{12} ,
\end{equation}
where $\Omega _{\alpha }$ represents rotational velocity of the frame with respect to the Fermi propagated one. 
In order to obtain a dimensionless dynamical system, we carry out expansion normalization. We adopt the following conventions:
\begin{equation}
\Sigma _{\pm } = \frac{1}{2H}(\sigma _{22}\pm \sigma _{33}), \ \ \ \ \ \Sigma _{\alpha \beta } = \frac{\sigma _{\alpha \beta }}{H} \ \ \ \ \ {\rm for} \ \ \ \ \ \alpha \neq \beta ,
\end{equation}
and
\begin{equation}
\fl \Phi = \frac{\dot{\phi}}{H} , \ \ \ \ \ \Psi = \frac{V_0 e^{-k\phi }}{3H^2 } ,\ \ \ \ \ \mathcal{E} = \frac{e^{\frac{g_1}{2}\phi }E}{\sqrt{6}H}, \ \ \ \ \ \mathcal{E}_{\alpha }^{(2)} = \frac{e^{\frac{g_2}{2}\phi }E_{\alpha }^{(2)}}{\sqrt{6}H} \ ,
 \quad \mathcal{E}_{\alpha }^{(A)} = \frac{e^{\frac{g_A}{2}\phi }E_{\alpha }^{(A)}}{\sqrt{6}H}  \ ,
\end{equation}
where capital Latins run from $3$ to $N$. The convenient choice of time coordinate $\tau $ is
\begin{equation}
d\tau = Hdt 
\end{equation}
and we use primes to denote the time derivatives.
The dynamical system is defined by the followings.
The evolution of geometry is governed by the following equations
\begin{eqnarray}
\fl \Sigma _{+} ^{\prime } = (q-2)\Sigma _{+} -(\Sigma _{12}^2 + \Sigma _{13}^2 ) +2\mathcal{E}^2 + 2(\mathcal{E}_1^{(2)})^2  - (\mathcal{E}_2^{(2)})^2 \nonumber  \\
  + \sum _{A = 3}^N (2(\mathcal{E}_1^{(A)} )^2 - (\mathcal{E}_2^{(A)} )^2 -(\mathcal{E}_3^{(A)})^2 ) \ ,  \\ \label{eq:start}
\fl \Sigma _{-}^{\prime } = (q-2)\Sigma _{-} - \Sigma _{12}^2 +\Sigma _{13}^2 +2\Sigma _{23}^2 -3(\mathcal{E}_2^{(2)})^2 -3 \sum _{A = 3}^N ((\mathcal{E}_2^{(A)})^2 - (\mathcal{E}_3^{(A)})^2 ) \ ,  \\
\fl \Sigma _{12} ^{\prime } = (q-2 +3\Sigma _{+}+\Sigma _{-})\Sigma _{12} +2\Sigma _{13}\Sigma _{23} -6\mathcal{E}_1^{(2)} \mathcal{E}_2^{(2)} -6 \sum _{A = 3}^N \mathcal{E}_1^{(A)} \mathcal{E}_2^{(A)} \ , \\
\fl \Sigma _{13} ^{\prime }  = (q-2+3\Sigma _{+}-\Sigma _{-})\Sigma _{13}  -6 \sum _{A = 3}^N \mathcal{E}_1^{(A)} \mathcal{E}_3^{(A)} \ , \\
\fl \Sigma _{23} ^{\prime } = (q-2-2\Sigma _{-})\Sigma _{23}-2 \Sigma _{12}\Sigma _{13}  -6 \sum _{A = 3}^N \mathcal{E}_2^{(A)} \mathcal{E}_3^{(A)} \ .
\end{eqnarray}
The equations for scalar field read
\begin{eqnarray}
\fl \Phi ^{\prime } = (q-2)\Phi +3k \Psi + 3g_1 \mathcal{E}^2+ 3g_2 \{ (\mathcal{E}_1^{(2)})^2 + (\mathcal{E}_2^{(2)})^2 \}
+3g_A \sum _{\alpha =1}^3  \sum _{A = 3}^N  ( \mathcal{E}_{\alpha }^{(A)})^2   \ ,\\
\Psi ^{\prime } = (2q+2-k\Phi ) \Psi \ .
\end{eqnarray}
The vector fields obey the following equations
\begin{eqnarray}
\mathcal{E}^{\prime } &=& (q-1-\frac{g_1}{2}\Phi -2\Sigma _{+})\mathcal{E}, \\
(\mathcal{E}_1^{(2)})^{\prime } &=& (q-1-\frac{g_2}{2}\Phi - 2\Sigma _{+})\mathcal{E}_1^{(2)} + 2\Sigma _{12}\mathcal{E}_2^{(2)} ,\\
(\mathcal{E}_2^{(2)})^{\prime } &=& (q-1-\frac{g_2}{2}\Phi +\Sigma _{+}+\Sigma _{-}) \mathcal{E}_2^{(2)} \ , 
\end{eqnarray}
and
\begin{eqnarray}
(\mathcal{E}_1^{(A)})^{\prime } &=& (q-1-\frac{g_A}{2}\Phi - 2\Sigma _{+})\mathcal{E}_1^{(A)} + 2\Sigma _{12}\mathcal{E}_2^{(A)}+2\Sigma _{13}\mathcal{E}_3^{(A)} , \\
(\mathcal{E}_2^{(A)})^{\prime } &=& (q-1-\frac{g_A}{2}\Phi +\Sigma _{+}+\Sigma _{-}) \mathcal{E}_2^{(A)} +2\Sigma _{23} \mathcal{E}_3^{(A)}, \\
(\mathcal{E}_3^{(A)} )^{\prime } &=& (q-1-\frac{g_A}{2}\Phi + \Sigma _{+}-\Sigma _{-})\mathcal{E}_3^{(A)} \ .
\end{eqnarray}
There is a constraint among the variables
\begin{eqnarray}
 1 &=& \Sigma _{+}^2 + \frac{1}{3}( \Sigma _{-}^2 + \Sigma _{12}^2 + \Sigma _{13}^2 + \Sigma _{23}^2 ) + \frac{1}{6}\Phi ^2 + \Psi \nonumber \\
&& + \mathcal{E}^2 +(\mathcal{E}_1^{(2)})^2 + (\mathcal{E}^{(2)}_2)^2  + \sum _{\alpha =1}^3  \sum _{A = 3}^N ( \mathcal{E}_{\alpha }^{(A)})^2 . 
\end{eqnarray}
The deceleration parameter $q= -1-\dot{H}/H^2 $ is given by
\begin{equation}
\fl q = 2\Sigma _{+}^2 + \frac{2}{3}(\Sigma _{-}^2 + \Sigma _{12}^2 + \Sigma _{13}^2 + \Sigma _{23}^2 ) + \frac{1}{3}\Phi ^2 - \Psi + \mathcal{E}^2 + (\mathcal{E}_1^{(2)})^2 + (\mathcal{E}^{(2)}_2)^2 + \sum _{\alpha =1}^3  \sum _{A = 3}^N  ( \mathcal{E}_{\alpha }^{(A)})^2   . \label{eq:end}
\end{equation}

Now, we are in a position to study the fate of anisotropy in the inflation with multiple vector
fields.

\section{Cases of uniform couplings $g\equiv g_m$}
\label{dynamical}

In this section, we restrict ourselves to the cases of uniform coupling constants $g\equiv g_m$
and show that isotropic inflation is an attractor
in the phase space for the number of vector fields greater than two. 
This clearly illustrates the existence of a principle replacing the cosmic no-hair conjecture.

\subsection{Multi-vector Isotropic Inflation}
The system of equations (\ref{eq:start}) to (\ref{eq:end}) forms a dynamical system of dimension $3(N+1)$. To understand the solution space, it is essential to find out equilibrium points and their linear stability.
Let us look for equilibrium points which have non-vanishing $\mathcal{E}$, $\mathcal{E}_2^{(2)}$ and $\mathcal{E}_3^{(A)}$. For their time derivatives to vanish, it is required that
\begin{equation}
q-1 = \frac{g}{2}\Phi, \ \ \ \ \ \Sigma _{+} = \Sigma _{-} = 0 .
\end{equation}
Then, from the evolution equations for $\mathcal{E}_1^{(2)}$ and $\mathcal{E}_{1,2}^{(A)} $, we obtain
\begin{equation}
\Sigma _{12}= \Sigma _{13}= \Sigma _{23} =0.
\end{equation}
Let us define the overall energy density parameter of the electric fields
\begin{equation}
\bar{\mathcal{E}}^2 = \mathcal{E}^2 + (\mathcal{E}_1^{(2)})^2 + (\mathcal{E}_2^{(2)})^2 + \sum _A (\mathcal{E}_{\alpha }^{(A)})^2 .
\end{equation}
Using the equilibrium conditions for $\Psi $ and $\Phi $ yeilds
\begin{equation}
q= \frac{k+g}{k-g} , \ \ \ \ \ \Phi = \frac{4}{k-g}, \ \ \ \ \ \bar{\mathcal{E}}^2 = \frac{k(k-g)-4}{(k-g)^2 } .
\end{equation}
Thus the existance condition for this class of solutions is $k(k-g) \geq 4$. 
In order to have an inflating (accelerating) universe, we also require $q<0 \Leftrightarrow k^2 -g^2 <0$.
The remaining task is to determine the relative strengths and angles among the field vectors through $\Sigma _{\alpha \beta } ^{\prime } =0$. There are six equations to satisfy for $3(N-1)$ variables.
\begin{eqnarray}
2\left( \mathcal{E}^2 +( \mathcal{E}_1^{(2)})^2 + \sum _A (\mathcal{E}_1^{(A)})^2 \right) = (\mathcal{E}_2^{(2)})^2 + \sum _A (\mathcal{E}_2^{(A)})^2 + \sum _A (\mathcal{E}_3^{(A)})^2, \\
(\mathcal{E}_2^{(2)})^2 + \sum _A (\mathcal{E}_2^{(A)})^2 = \sum _A (\mathcal{E}_3^{(A)})^2 , \\
\mathcal{E}_1^{(2)} \mathcal{E}_2^{(2)} + \sum _A \mathcal{E}_1^{(A)} \mathcal{E}_2^{(A)} = 0 , \\
\sum _A \mathcal{E}_1^{(A)} \mathcal{E}_3^{(A)}  = 0 , \\
\sum _A \mathcal{E}_2^{(A)} \mathcal{E}_3^{(A)} = 0 , \\
 \mathcal{E}^2 + (\mathcal{E}_1^{(2)})^2 + (\mathcal{E}_2^{(2)})^2 + \sum _{A, \alpha } (\mathcal{E}_{\alpha }^{(A)})^2 = \frac{k(k-g)-4}{(k-g)^2 } .
\end{eqnarray}
When $N=3$, they lead to a unique orthogonal solution
\begin{equation}
\mathcal{E}^2 = (\mathcal{E}_2^{(2)})^2 = (\mathcal{E}_3^{(3)})^2 = \frac{k(k-g)-4}{3(k-g)^2 },\ \ \ \ \ \mathcal{E}_1^{(2)} = \mathcal{E}_1^{(3)} = \mathcal{E}_2^{(3)} = 0 .
\end{equation}
For $N>3$, it is convenient to introduce $N-2$ dimensional vector notation
\begin{equation}
\vec{\mathcal{E}}_{\alpha } = \left(   \begin{array}{c}
    \mathcal{E}_{\alpha }^{(3)} \\ 
     \mathcal{E}_{\alpha }^{(4)} \\ 
     \vdots \\ 
    \mathcal{E}_{\alpha }^{(N)}  \\ 
  \end{array} \right).
\end{equation}
The magnitude of $\vec{\mathcal{E}}_3 $ is given by
\begin{equation}
|\vec{\mathcal{E}}_3|^2 =  \frac{k(k-g)-4}{3(k-g)^2 } .
\end{equation}
The magnitude of $\vec{\mathcal{E}}_2$ is controled by $\mathcal{E}_2^{(2)}$ through
\begin{equation}
(\mathcal{E}_2^{(2)})^2 + |\vec{\mathcal{E}}_2|^2 =  \frac{k(k-g)-4}{3(k-g)^2 } .
\end{equation}
Similarly, we have, for $\vec{\mathcal{E}}_1$,
\begin{equation}
\mathcal{E}^2 + (\mathcal{E}_1^{(2)})^2 + | \vec{\mathcal{E}}_1|^2=  \frac{k(k-g)-4}{3(k-g)^2 } .
\end{equation}
They introduce three arbitrary constant parameters. We also know from
\begin{equation}
\vec{\mathcal{E}}_1 \cdot \vec{\mathcal{E}}_3 = \vec{\mathcal{E}}_2 \cdot \vec{\mathcal{E}}_3 = 0
\end{equation}
that $\vec{\mathcal{E}}_3$ is perpendicular to $\vec{\mathcal{E}}_1$ and $\vec{\mathcal{E}}_2$. The angle between $\vec{\mathcal{E}}_1$ and $\vec{\mathcal{E}}_2$ is fixed once we choose the three parameters controling their magnitudes by using
\begin{equation}
\mathcal{E}_1^{(2)} \mathcal{E}_2^{(2)} + \vec{\mathcal{E}}_1 \cdot \vec{\mathcal{E}}_2 = 0.
\end{equation}
Apart from these geometrical conditions, we can take arbitrary combinations of the componenets for these three ($N-2$ dimensional) vectors. In particular, there is $O(N-2)$ invariance for the equilibrium values of $\mathcal{E}_{\alpha }^{(A)}$. Therefore, they are a family of equilibrium points that spans a $3(N-3)$ dimensional submanifold in the state space, for which the spatial slices are isotropic even though the vector fields have non-zero background values.

\subsection{Stability of the Isotropic Inflation}

To decide local stability of an equilibrium point requires linearized equations around it. Let us pick up one of the isotropic equilibrium points derived above. Before linearization, we shall manipulate the variables so that the stability analysis around that particular point simplifies. First, apply a rotation $(R)^{(A)}_{\ (B)}$ in the $N-2$ dimensional internal space to introduce new variables $\mathcal{D}_{\alpha }^{(A)} = (R)^{(A)}_{\ (B)} \mathcal{E}_{\alpha }^{(B)}$ such that their values at that equilibrium satisfy
\begin{equation}
 \vec{\mathcal{D}}_1 = \left( \begin{array}{c} \mathcal{D}_1^{(3)} \\ \mathcal{D}_1^{(4)} \\ 0 \\ \vdots \\ 0
\end{array} \right), 
\ \ \ \ \  \vec{\mathcal{D}}_2 = \left( \begin{array}{c} \mathcal{D}_2^{(3)} \\ \mathcal{D}_2^{(4)} \\ 0 \\ \vdots \\ 0
\end{array} \right), 
\ \ \ \ \ \vec{\mathcal{D}}_3 = \left( \begin{array}{c} 0 \\ 0 \\ \mathcal{D}_3^{(5)} \\ 0 \\ \vdots \\ 0
\end{array}\right) \neq 0 .
\end{equation}
This operation leaves the equations unchanged. When liearized around that equilibrium point, the perturbations for $\mathcal{D}_{\alpha }^{(B)}$ for $B\geq 6$ as well as $\mathcal{D}_3^{(3)}$ and $\mathcal{D}_3^{(4)}$ give trivially zero-eigenvalues. Next, notice that at the equilibrium
\begin{equation}
\mathcal{E}_1^{(2)} \mathcal{E}_2^{(2)} + \mathcal{D}_1^{(3)}\mathcal{D}_2^{(3)} + \mathcal{D}_1^{(4)}\mathcal{D}_2^{(4)} =0 .
\end{equation}
Thus we can choose a $3$ by $3$ orthogonal matrix $Q$ such that
\begin{equation}
\fl \left( \begin{array}{c} \mathcal{C}_1^{(2)} \\ \mathcal{C}_1^{(3)} \\ \mathcal{C}_1^{(4)} \end{array} \right) 
= Q \left( \begin{array}{c} \mathcal{E}_1^{(2)} \\ \mathcal{D}_1^{(3)} \\ \mathcal{D}_1^{(4)} \end{array}\right) = \left( \begin{array}{c} \mathcal{C}_1^{(2)} \\ 0\\ 0 \end{array} \right) , \ \ \ \ \ 
 \left( \begin{array}{c} \mathcal{C}_2^{(2)} \\ \mathcal{C}_2^{(3)} \\ \mathcal{C}_2^{(4)} \end{array}\right) 
= Q \left( \begin{array}{c} \mathcal{E}_2^{(2)} \\ \mathcal{D}_2^{(3)} \\ \mathcal{D}_2^{(4)} \end{array}\right) = \left( \begin{array}{c}  0 \\ \mathcal{C}_2^{(3)} \\ 0 \end{array} \right),
\end{equation}
where the last equalities for each vector hold for the equilibrium. The linearized equations are invariant under this change and the eigenvalues for $\mathcal{C}_1^{(4)}$, $\mathcal{C}_2^{(4)}$ and $\mathcal{C}_2^{(2)}$ are zero. Finally we take a $2$ by $2$ orthogonal matrix $P$ to introduce a new pair of variables
\begin{equation}
\left( \begin{array}{c} \mathcal{E}^{(1)} \\ \mathcal{E}^{(2)} \end{array}\right) = P\left( \begin{array}{c}\mathcal{E} \\ \mathcal{C}^{(2)}_1 \end{array}\right)
\end{equation}
such that $\mathcal{E}^{(2)} =0$ is satisfied at the equilibrium and its eigenvalue there is zero. After these transformations, we are left with $\mathcal{E}^{(1)}$, $\mathcal{C}_1^{(3)}$, $\mathcal{C}_2^{(3)}$ and $\mathcal{D}_{\alpha }^{(5)}$ whose eigenvalues are potentially non-zero. All the other zero-eigenvalues are manifestation of the fact that we are considering a hypersurface of equilibrium points. Looking at the resulting linearized equations, we see that the problem is now reduced to the stability analysis of the orthogonal equilibrium point with three electric fields. The correspondance to the notation for $N=3$ is
\begin{equation}
\mathcal{E}^{(1)} \leftrightarrow \mathcal{E}, \ \ \ \ \ \mathcal{C}_{1,2}^{(3)} \leftrightarrow \mathcal{E}_{1,2}^{(2)}, \ \ \ \ \ \mathcal{D}_{\alpha }^{(5)} \leftrightarrow \mathcal{E}_{\alpha }^{(3)}.
\end{equation}

The stability analysis for $N=3$ is straightforward. The twelve independent varibales ($\Psi $ is eliminated by using the Hamiltonian constraint) are grouped into six pairs: $\Sigma _{12}$ - $\mathcal{E}_1^{(2)}$, $\Sigma _{13}$ - $\mathcal{E}_1^{(3)}$, $\Sigma _{23}$ - $\mathcal{E}_2^{(3)}$, $\Sigma _{-}$ - $\mathcal{E}_2^{(2)} - \mathcal{E}_3^{(3)}$, $\Sigma _{+}$ - $2\mathcal{E}-\mathcal{E}_2^{(2)}-\mathcal{E}_3^{(3)}$ and $\Phi $ - $\mathcal{E}+\mathcal{E}_2^{(2)}+ \mathcal{E}_3^{(3)}$. The first five pairs, which represent anisotropic perturbations,
 share common eigenvalues given by
\begin{equation}
\lambda _{\Sigma \ \mathcal{E}}= \frac{q-2 \pm \sqrt{(q-2)^2 - 16\bar{\mathcal{E}}^2}}{2}.
\end{equation}
Their real parts are negative since $-1 \leq q \leq 2$. 
 The isotropic mode has
\begin{equation}
\lambda _{\Phi \ \mathcal{E}} = \frac{q-2 \pm\sqrt{(q-2)^2 - 4 \bar{\mathcal{E}}^2 (4-3g(k-g))}}{2} .
\end{equation}
If we require $q = \frac{k+g}{k-g}<0$, it follows $4-3g(k-g)>0$, which means that both of the eigenvalues have negative
real part again. 
 In other word, these isotropic equilibrium points are local sinks for a range of the parameters for which the universe undergoes rapidly accelerated expansion.

One important consequence of this result (together with the global analysis in the next section) 
is that the anisotropy due to non-Abelian gauge fields
should disappear if duration of inflation is sufficiently long since they can generically be regarded 
as a number of copies of Maxwell vector field, interacting with each other. 
The sufficient condition for isotropic final state is weakness of the Yang-Mills gauge couplings during inflation, 
which appears to be satisfied in the gauge-kinetic coupling model \cite{Murata:2011wv}.

\section{The Cosmic Minimum-Hair Conjecture}

In this section, we would like to gather evidences for the cosmic minimum-hair conjecture which
we would like to propose. In the first part, we clarify the global dynamical structure of the uniform coupling models.
There, we see a mechanism of isotropization due to the generic structure of Einstein-Maxell equations.
In the second part, we try to explore the fate of the anisotropy in models with more general couplings. 
We will see that the anisotropy of the universe is monotonically decreasing.

\subsection{Global phase space structure}

A natural question arises; whether those isotropic inflations with non-vanishing vector fields are the only attractor of the system. It is difficult to derive a global conclusion for a system of this complexity in general. It is also a parameter dependent problem. However, we can argue that multiple vector fields are expected to repel each other and try to become isotropic in expanding universes. 

Let us first look at the stability of axisymmetric inflating solutions which were discovered in \cite{Kanno:2010nr}. In the present model, they are located on lower dimensional boundaries of the full state space. For example, we have an equilibirum point
 \begin{equation*}
\fl q = \frac{5k^2-2kg -3g^2 -8}{(k-3g)(k-g)+8}, \ \ \ \ \ \ \Sigma _{+} = \frac{2(k^2 -kg -4)}{(k-3g)(k-g)+8}, \ \ \ \ \ \Phi = \frac{12(k-g)}{(k-3g)(k-g)+8}, 
\end{equation*}
\begin{equation*}
\fl \Psi = \frac{3((k+3g)(k-g)-8)(g(k-g)-4)}{[(k-3g)(k-g)+8]^2 }, \ \ \ \ \ \mathcal{E}^2  = -\frac{3(k(k-g)-4)((k+3g)(k-g)-8)}{[(k-3g)(k-g)+8]^2 }, 
\end{equation*}
\begin{equation*}
 \Sigma _{-} = \Sigma _{12} = \Sigma _{13} = \Sigma _{23} = \mathcal{E}_{1,2}^{(2)} = \mathcal{E}_{\alpha }^{(A)} =  0 .
\end{equation*}
There are many others which represent physically the same spacetime, but lie on different boundaries. It was a stable attractor solution for the single-vector-field model. For this equilibrium state, we notice that $\Sigma _{+} \geq 0$ by looking at the evolution equation for $\Sigma _{+}$ and requiring $\mathcal{E}^2 \geq 0$.  More generally, any electric field in 1-direction tends to support positive $\Sigma _{+}$ by generating a tension along that direction. It is also easy to see that for any electric perturbation in 2- or 3-direction, the eigenvalue is given by $3\Sigma _{+} \geq 0$. Thus this axisymmetric solution is a saddle point in the general multi-field state space. Putting these together, the mathematical structure of the Maxwell's equations is such that the positive $\Sigma _{+}$ created by the 1-component of a field comes with negative sign in the eveolution equation in 1-direction while it has plus signs in the other directions. Thus, if there are more than one vector fields and one of them is dominant, it creates such an anistoropy that destabilizes orthogonal components of the other fields. This instability does not show up for single field models since it would merely cause a rotation of that vector. If those orthogonal components grow and surpass the original vector, it then becomes unstable and recovers its amplitude.

Thus we expect that multiple vector fields always try to rearrange their orientations so as to minimize anisotropy of the space which would cause an instability in some of their components. This is a generic feature of the Maxwell's equations in general relativity and has nothing to do with any couplings to other matter components. The point is that the anisotropy created by a tension acts rather to reduce the tension itself. In the present model, the vectors cannot disappear since they keep excited by the dominant scalar field. However, they still redistribute themselves to achieve as much isotropy as possible within the given circumstance.  As an example, if $N=2$, the attractor solution contains two orthogonal electric fields with the same amplitude:
\begin{equation*}
\fl q = \frac{2k^2 -2kg -3g^2 -2}{(k-3g)(k-g)+2}, \ \ \ \ \ \Sigma _{+}= \frac{k(k-g)-4}{2(k-3g)(k-g)+4}, \ \ \ \ \ \Sigma _{-} = - \frac{3k(k-g)-12}{2(k-3g)(k-g)+4} ,
\end{equation*}
\begin{equation*}
 \Phi = \frac{6(k-2g)}{(k-3g)(k-g)-2}, \ \ \ \ \ \Psi = \frac{3(g(k-g)-2)(g(2k-3g)-2)}{((k-3g)(k-g)+2)^2}, 
\end{equation*}
\begin{equation*}
\mathcal{E}^2 = (\mathcal{E}_2^{(2)})^2 = -\frac{3(k(k-g)-4)(g(2k-3g)-2)}{2((k-3g)(k-g)+2)^2} ,
\end{equation*}
\begin{equation*}
\Sigma _{12}= \Sigma _{13}=\Sigma _{23} = \mathcal{E}_1^{(2)} = 0 .
\end{equation*}
If we include a third vector field, however, it becomes unstable because of the eigenvalue for $\mathcal{E}_3^{(3)}$
\begin{equation}
\lambda _{\mathcal{E}_3} = 6\Sigma _{+} >0 .
\end{equation}

From the above argument, we can also read off the global behaviour of the orbits. There are a bunch of saddle points consisting of those axisymmetric signle-field and orthogonal two-field equilibrium solutions on the boundaries, which are attractors when restricted in properly chosen invariant subsystems. Since these invariant sets have dimensions smaller than $3(N+1)$, the initial condition from which the orbits are attracted to any of the anisotropic attractors is of measure zero in the entire
state space. Whenever an orbit starts from a point not included in these subsystems, it is attracted 
to the isotropic final state after a sufficiently long time. For example, in order for an orbit to be attracted 
to the two field attractor, it must satisfy $\mathcal{E}_{\alpha }^{(A)}=0$ for all $A=3,\cdots ,N$. Any small deviation from this subsystem
would lead it to isotropy due to the linear instability demonstrated above. A typical orbit is first attracted towards a nearby saddle point. Then the instability explained above kicks in and another vector component rises. It might come across another instability and go to another saddle point. An orbit continues this routine until it finally settles down to one of the isotropic attractors.

\subsection{The cosmic minimum-hair conjecture}

In this subsection, we numerically solve the dynamical equations (\ref{eq:start}) - (\ref{eq:end}) in the case of three vector fields.
The results show that the cosmic minimum-hair conjecture indeed holds in many cases. 

First of all, we confirm the dynamical mechanism discussed in the previous section. 
We solved the basic equations with the parameters $k= -2$, $g=5$, 
and the initial conditions  $\Sigma_\pm = \Sigma_{12} = \Sigma_{13} = \Sigma_{23} =0$,
$\Phi =0.6 , \mathcal{E} = 0.1, \mathcal{E}^{(2)}_1 = 0.2 \ ,  \mathcal{E}^{(2)}_2  = 0.001 \ , 
\mathcal{E}^{(3)}_1  = 0.1 \ , \mathcal{E}^{(2)}_2  = 0.01 \ , \mathcal{E}^{(2)}_3  = 0.0001 $. 
The initial condition for  $\Psi$ is determined by the constraint equation.
The parameters are chosen so that the universe is accelerating, namely $q <0$.

 \begin{figure}[ht]
 \begin{center}
\includegraphics[height=6cm, width=7.5cm]{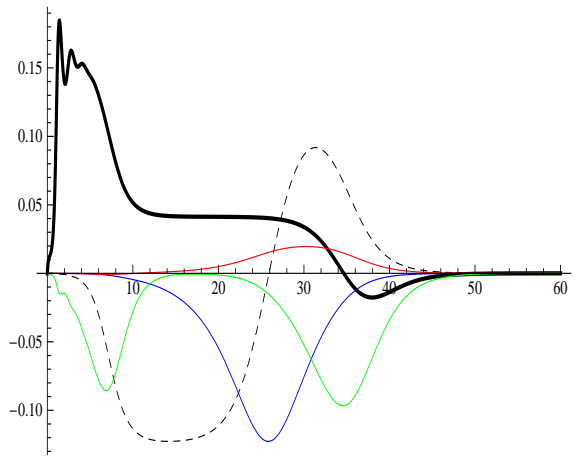}
\end{center}
\caption{For a uniform coupling $g=5$, we plotted time evolutions of anisotropic expansion rate normalized by the Hubble.
The horizontal axis is the e-folding number.
The components $\Sigma_+$,  $\Sigma_-$,  $\Sigma_{23}$,
$\Sigma_{13}$, and  $\Sigma_{12}$ correspond to thick, dashed, blue, red, and green lines, respectively.}
\label{fg:sigma}
\end{figure}

From Fig.\ref{fg:sigma}, we see the anisotropy disappears after transient anisotropic inflationary phases.
This confirms that the isotropic inflation is an attractor. The linear stability of the isotropic solutions 
shown in the previous subsection guarantees that the final isotropic plateau is not transient 
and never decays. It should be noted that 
the vector fields possess a nontrivial orthogonal configuration.
We have solved the basic equations for other sets of parameters and initial conditions.
Although the transient behavior depends on the parameters and the initial conditions,
the system always achieves the isotropic inflationary final state asymptotically. 
The transient anisotropic phases correspond to the saddle points discussed in the previous section.
However, since duration of anisotropic inflation is sufficiently long, the anisotropy at each of the saddle points
would be relevant to CMB observations in the realistic cases where inflation ends with a finite duration.

 \begin{figure}[ht]
 \begin{center}
\includegraphics[height=6cm, width=7.5cm]{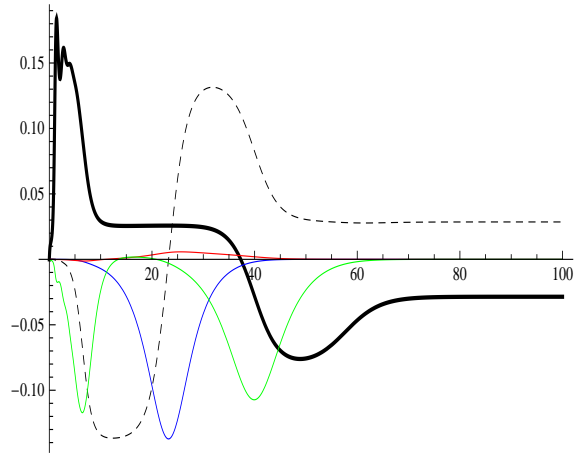}
\end{center}
\caption{The time evolution for the couplings $g_1=4.8 \ , g_2=5.0 \ , g_3=5.2$. The horizontal axis is the e-folding number.
The components $\Sigma_+$,  $\Sigma_-$,  $\Sigma_{23}$,
$\Sigma_{13}$, and  $\Sigma_{12}$ correspond to thick, dashed, blue, red, and green lines, respectively.}
\label{fg:sigma-a}
\end{figure}
 \begin{figure}[ht]
 \begin{center}
\includegraphics[height=6cm, width=7.5cm]{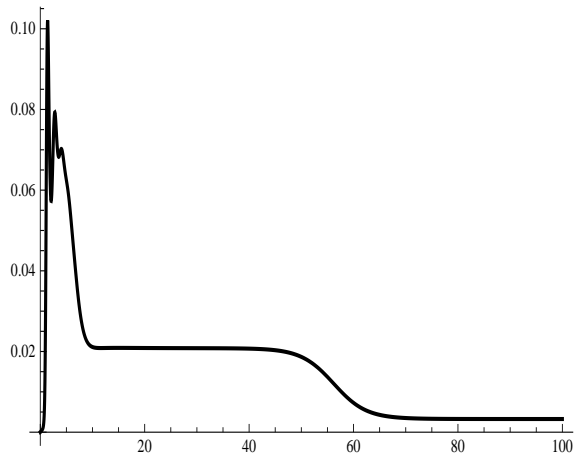}
\end{center}
\caption{The time evolution of the total anisotropy $\Sigma^{ab} \Sigma_{ab} /2$ is depicted with respect to
the e-folding number.}
\label{fg:total}
\end{figure}

In order to see what happens for general couplings, we took couplings
$g_1= 4.8 \ , g_2= 5.0 \ , g_3 = 5.2 $ for the three vector fields.
As you can see in Fig.\ref{fg:sigma-a}, the attractor is an anisotropic inflation. Strictly speaking, 
the final plateau in Fig.\ref{fg:sigma-a} might decay eventually if we waited long enough since 
we didn't prove the existence of a stable fixed point for nonuniform couplings, although 
numerical evidence suggests otherwise. 
In this general case, the behavior is even more interesting. Since there are several saddle points, 
we  have different anisotropic inflationary stages. During inflation the preferred direction in the universe is changing, which would lead to interesting observational consequences. For example, the anisotropy
can be scale dependent.  Even if we change initial conditions, the attractor is the same.
Hence, the predictive power of the theory is maintained.
In Fig.\ref{fg:total}, we plotted the square of the magnitude of anisotropy $\Sigma^{ab} \Sigma_{ab} /2$. 
While the system is not close to any of the fixed points, the orbit follows a complicated dynamical trajectory.
However, once it starts to be attracted towards one of the inflating solutions, the evolution
is fairly regular. 
The result shows monotonic decrease of the total anisotropy after the orbit goes into this regular 
tracking regime in spite that each component of the anisotropy exhibits a complicated behavior. 
It is inferred that when an orbit goes from a fixed point to another, the latter is always less anisotropic
than the former. We have examined a lot of other parameters and initial conditions and
confirmed that the total anisotropy is always monotonically decreasing at late times.

The above evidence strongly supports the cosmic minimum-hair conjecture: that is, the universe organizes itself
so that any feature of the spacetime during inflation becomes minimum. The cosmic no-hair theorem can be regarded as a special case of
this general principle.

\section{Conclusion}

We investigated inflationary models with multiple vector fields. In the presence of non-trivial couplings
between the inflaton and the vector fields, the no-hair theorem does not hold any more.
The universe can possess vectorial hairs in general. 
Thus, for a single or two vector fields, there remains anisotropy of the expansion.
Remarkably, the expansion of the universe eventually becomes isotropic 
when the number of vector fields is greater than two and the couplings are uniform.
More precisely, while one or two vector fields dominate the energy density of the vector sector,
the expansion of the universe is anisotropic. However, since these phases are transient in the presence of other vectors,
the trajectory eventually approaches the attractor, which has been shown to be the isotropic multi-vector inflation.
This isotropization mechanism tells us that orthogonal triad is a stable configuration.
This also tells us non-Abelian gauge fields do not create any anisotropy in the asymptotic future unless they remain in a strong-coupling regime during inflation.
Curiously, a similar mechanism works in a different vector inflation model~\cite{Maleknejad:2011jr}.
In view of these results, one may think that the no-hair conjecture survives. 
However, it should be stressed that there exists vectorial hair.
The point is that vector fields isotropize the universe by themselves. In the case of many vector fields,
the configuration cannot be an orthogonal system, rather 
the eventual configuration for vector fields depends on the initial conditions. 
In a sense, this is a ``moduli space" for isotropic inflation with multi-vector fields. 
When the coupling constant $g_m$ for each vector field is different, 
we have anisotropic inflation as the attractor. Even in this case, there are several transient anisotropic inflationary phases.
It seems that the inflating system dynamically selects a minimally anisotropic configuration.
This has been confirmed numerically. 
On the ground of these analyses, we can expect that the cosmic minimum-hair conjecture holds.

There is a caveat. In this paper, we have considered a power law type inflation, which is eternal.
However, in the realistic inflationary models there exists the end of the inflation. Although the 
analytical treatment in the present article does not work for general scalar potential, the system
would still be well-behaved and numerical calculation should be readily implemented. 
To study the end of inflation will be an interesting direction for further research not least 
because of its importance regarding reheating.
We would observe anisotropy if the saddle solution earns a sufficient e-folding number and
the inflation ends before the relevant scales go beyond the observable range. 
The phenomenological consequence of the anisotropy is worth being mentioned. 
It produces the statistical anisotropy and the cross correlation between curvature perturbations and
primordial gravitational waves~\cite{Himmetoglu:2009mk,Dulaney:2010sq,Gumrukcuoglu:2010yc,Watanabe:2010fh}. 
It would be interesting to see these features in the CMB~\cite{Watanabe:2010bu,Shiraishi:2011ph}.

\ack
 JS would like to thank M. M. Sheikh-Jabbari, Sugumi Kanno, Azadeh Maleknejad and Shinji Mukohyama for useful discussion
on the cosmic no-hair conjecture. KY would like to thank Keiju Murata for advice and encouragement. KY is supported by Cambridge Overseas Trust. 
This work is partially supported by  the
Grant-in-Aid for  Scientific Research Fund of the Ministry of 
Education, Science and Culture of Japan (C) No.22540274, (A) (No. 21244033, No.22244030), the
Grant-in-Aid for  Scientific Research on Innovative Area No.21111006,
JSPS under the Japan-Russia Research Cooperative Program
and the Grant-in-Aid for the Global COE Program 
``The Next Generation of Physics, Spun from Universality and Emergence".

\section*{References}


\begin{thebibliography}{99}

\bibitem{Komatsu:2010fb}
  Komatsu E {\it et al.} 2001 Seven-Year Wilkinson Microwave Anisotropy Probe (WMAP) Observations: Cosmological Interpretation {\it Preprint} arXiv:1001.4538 [astro-ph.CO]

\bibitem{Wald:1983ky}
Wald R M 1983 {\it Phys. Rev. D} {\bf 28} 2118


\bibitem{varun}
Moss I and Sahni V 1986 {\it Phys. Lett. B} {\bf 178}  159

\bibitem{Ford:1989me}
Ford L H 1989 {\it Phys. Rev. D} {\bf 40} 967

\bibitem{Kaloper:1991rw}
  Kaloper N 1991 {\it Phys. Rev. D} {\bf 44} 2380

  
\bibitem{Barrow:2005qv}
 Barrow J D and Hervik S 2006 {\it Phys. Rev. D} {\bf 73} 023007


\bibitem{Barrow:2009gx}
Barrow J D and Hervik S 2010 {\it Phys. Rev. D} {\bf 81} 023513

  
\bibitem{Campanelli:2009tk}
Campanelli L 2009 {\it Phys. Rev. D} {\bf 80} 063006

  
\bibitem{Golovnev:2008cf}
Golovnev A, Mukhanov V and Vanchurin V 2008 {\it J. Cosmo. Astropart. P.} {\bf 0806} 009

  
\bibitem{Kanno:2008gn}
Kanno S, Kimura M, Soda J and Yokoyama S 2008 {\it J. Cosmo. Astropart. P.} {\bf 0808} 034


\bibitem{Ackerman:2007nb}
Ackerman L, Carroll S M and Wise M B 2007 {\it Phys. Rev. D} {\bf 75} 083502

   
\bibitem{Himmetoglu:2008zp}
  Himmetoglu B, Contaldi C R and Peloso M 2008 Instability of anisotropic cosmological solutions supported by vector fields {\it Preprint} arXiv:0809.2779 [astro-ph];
  Himmetoglu B, Contaldi C R and Peloso M 2008 Instability of the ACW model, and problems with massive vectors during inflation {\it Preprint} arXiv:0812.1231 [astro-ph];
  Himmetoglu B, Contaldi C R and Peloso M 2009 {\it Phys. Rev. D} {\bf 80} 123530

\bibitem{Golovnev:2009rm} 
  Golovnev A 2009 {\it Phys. Rev. D} {\bf 81} 023514

\bibitem{EspositoFarese:2009aj} 
  Esposito-Farese G, Pitrou C and Uzan J P 2010 {\it Phys. Rev. D} {\bf 81} 063519

\bibitem{Watanabe:2009ct}
  Watanabe M a, Kanno S and Soda J 2009 {\it Phys. Rev. Lett.} {\bf 102} 191302

\bibitem{Kanno:2009ei}
  Kanno S, Soda J and Watanabe M a 2009 {\it J. Cosmo. Astropart. P.} {\bf 0912} 009

\bibitem{Kanno:2010nr} 
  Kanno S, Soda J and Watanabe M -a 2010 {\it J. Cosmo. Astropart. P.} {\bf 1012} 024

\bibitem{Moniz:2010cm}
Moniz P V and Ward J 2010 Gauge field back-reaction in Born Infeld cosmologies {\it Preprint} arXiv:1007.3299 [gr-qc]
 
\bibitem{Hassan}
  Emami R, Firouzjahi H, Movahed S M S and Zarei M 2010 Anisotropic Inflation from Charged Scalar Fields {\it Preprint} arXiv:1010.5495 [astro-ph.CO]

\bibitem{Dimopoulos:2010xq} 
  Wagstaff J M and Dimopoulos K 2011 {\it Phys. Rev. D} {\bf 83} 023523

\bibitem{Do:2011zz} 
  Do T Q, Kao W F and Lin I C 2011 {\it Phys. Rev. D} {\bf 83} 123002

\bibitem{Murata:2011wv} 
  Murata K and Soda J 2011 {\it J. Cosmo. Astropart. P.} {\bf 1106} 037

\bibitem{Bhowmick:2011em} 
  Bhowmick S and Mukherji S 2011 Anisotropic Power Law Inflation from Rolling Tachyons {\it Preprint}  arXiv:1105.4455 [hep-th]

\bibitem{Hervik:2011xm} 
  Hervik S, Mota D F and Thorsrud M 2011 {\it J. High Energy Phys.} {\bf 1111} 146
  
\bibitem{Himmetoglu:2009mk}
  Himmetoglu B 2009 Spectrum of Perturbations in Anisotropic Inflationary Universe with Vector Hair {\it Preprint} arXiv:0910.3235 [astro-ph.CO]

\bibitem{Dulaney:2010sq}
  Dulaney T R and Gresham M I 2010 Primordial Power Spectra from Anisotropic Inflation {\it Preprint} arXiv:1001.2301 [astro-ph.CO]

  
\bibitem{Gumrukcuoglu:2010yc}
  Gumrukcuoglu A E, Himmetoglu B and Peloso M 2010 Scalar-Scalar, Scalar-Tensor, and Tensor-Tensor Correlators from Anisotropic Inflation {\it Preprint} arXiv:1001.4088 [astro-ph.CO]
  
\bibitem{Watanabe:2010fh}
  Watanabe M a, Kanno S and Soda J 2010 {\it Prog. Theor. Phys.} {\bf 123} 1041

\bibitem{Watanabe:2010bu}
Watanabe M a, Kanno S and Soda J 2010 Imprints of Anisotropic Inflation on the CMB {\it Preprint} arXiv:1011.3604 [astro-ph.CO]
  

\bibitem{Ellis:1968vb} 
 Ellis G F R and MacCallum M A H 1969 {\it Commun. Math. Phys.} {\bf 12} 108

\bibitem{WE}
Wainwright J and Ellis G F R (Eds.) 1997 {\it Dynamical Systems in Cosmology}
 (Cambridge: Cambridge University Press)
 
\bibitem{Maleknejad:2011jr} 
  Maleknejad A, Sheikh-Jabbari M M and Soda J 2011 Gauge-flation and Cosmic No-Hair Conjecture {\it Preprint} arXiv:1109.5573 [hep-th]

\bibitem{Shiraishi:2011ph} 
  Shiraishi M and Yokoyama S 2011 {\it Prog. Theor. Phys.} {\bf 126} 923

\end{thebibliography}
\end{document}